\documentclass[10pt,preprint]{aastex}
\usepackage{graphicx}
\usepackage{soul}
\usepackage{color}

\shorttitle{Asymmetric transit curves as indication of orbital obliquity: KOI-13}
\shortauthors{Szab\'o et al.}

\begin{document}

\title{Asymmetric transit curves as indication of orbital obliquity: clues from the late-type dwarf companion in KOI-13}

\author{
Gy. M. Szab\'o\altaffilmark{1,2},
R. Szab\'o\altaffilmark{1},
J. M. Benk\H{o}\altaffilmark{1}, 
H. Lehmann\altaffilmark{3}, 
Gy. Mez\H{o}\altaffilmark{1}, 
A. E. Simon\altaffilmark{1}, 
Zs. K\H{o}v\'ari\altaffilmark{1}, 
G. Hodos\'an\altaffilmark{1,4},
Zs. Reg\'aly\altaffilmark{1},
L. L. Kiss\altaffilmark{1,5}
}
\altaffiltext{1}{Konkoly Observatory of the Hungarian Academy of Sciences, PO. Box 67, H-1525 Budapest, Hungary}
\altaffiltext{2}{Dept. of Exp. Phys., University of Szeged, Hungary}
\altaffiltext{3}{Th\"uringer Landessternwarte, 07778 Tautenburg, Germany}
\altaffiltext{4}{Department of Astronomy, E\"otv\"os University, P\'azm\'any P\'eter s\'et\'any 1/A, 1117 Budapest, Hungary }
\altaffiltext{5}{Sydney Institute for Astronomy, School of Physics A28, University of Sydney, NSW 2006, Australia}

\begin{abstract}
KOI-13.01, a planet-sized companion in an optical double star was announced as one of the 1235 $Kepler$ planet candidates in February 2011. The transit curves show significant distortion that was stable over the $\sim$130 days time-span of the data. Here we investigate the phenomenon via detailed analyses of the two components of the double star and a re-reduction of the $Kepler$ data with pixel-level photometry. Our results indicate that KOI-13 is a common proper motion  binary, with two rapidly rotating components ($v\sin i\approx$65--70~km/s). We identify the host star of KOI-13.01 and conclude that the transit curve asymmetry is consistent with a companion orbiting a rapidly rotating, possibly elongated star on an oblique orbit. The radius of the transiter is 2.2~$R_J$, implying an {irradiated late-type dwarf, probably a} hot brown dwarf rather than a planet. KOI-13 is the first example for detecting orbital obliquity for a substellar companion without measuring the Rossiter-McLaughlin effect with spectroscopy.
\end{abstract}

\keywords {stars: binaries: individual: CCDM J19079+4652AB--- planetary systems: individual: KOI-13 --- brown dwarfs: individual: KOI-13.01}

\section{Introduction}
Barnes (2009) suggested that exoplanets orbiting rapidly rotating
stars may have unusual light curve shapes. These objects transit across an oblate disk with non-isotropic surface brightness, caused by the gravitational darkening. The von Zeipel (1924) theorem predicts that the flux emitted from the surface of a rapidly rotating star is proportional to the local effective gravity. Therefore, rapid rotation of early-type stars induce an equator-to-pole gradient in the effective temperature. This prediction was recently confirmed by interferometric observations (e.g. Ciardi et al. 2001, Peterson et al. 2006), however, standard models of uniformly rotating stars cannot predict the quantitative findings (Monnier et al. 2007, Che et al. 2011). If a substellar object transits such a star, unusual and distinctive transit light curves are expected. If such asymmetries are measured, one can detect the orbital obliquity of the exoplanet, even without the analysis of the Rossiter-McLaughlin effect in the spectrum (Rossiter 1924, McLaughlin 1924, Gaudi and Winn 2007). 

Winn et al. (2010) demonstrated the preference of oblique orbits for hot Jupiters orbiting hot stars ($T_{\rm eff}>6250$~K). They suggested two alternative hypotheses: (i) the production of hot Jupiters is bimodal: low-mass stars produce a lower number of hot Jupiters which spiral-in the protoplanetary disk, while massive stars form more of massive planets in unstable orbits which then suffer vivid scattering; (ii) there is a single pathway of production of hot Jupiters, but during later evolutionary stages, the interaction between the planet and the convective zone of the star is insignificant for $T_{\rm eff}>6250$~K, which results in the observed bimodality. Currently, there is not enough information on the obliquity distribution of hot brown dwarfs, although one would expect a similar behavior for brown dwarfs and exoplanets in this sense. The discovery of many companions in oblique orbits and the characterization of the host stars is fundamental to choose between the proposed scenarios.

Borucki et al. (2011) published light curves of 1235 planetary candidates with transit-like signatures detected by the $Kepler$ satellite mission between 2 May -- 16 September 2009. About 2\%{} of host stars have $T_{\rm eff}>6500 K$, which makes this data set suitable to search for the impact of gravity darkening effect on the measured light curves.
We scanned all data, searching for asymmetric transit light curves of early-type stars, and found KOI-13 to be a prime example for such an object. Borucki et al. (2011) remarked that KOI-13 is double, thus a sophisticated analysis was necessary, including follow-up observations. In this paper, we analyse the distorted light curve of KOI-13 (Sect. 2), give evidence for the binary nature of KOI-13 and characterize the components (Sect. 3). Finally, we identify the host star of the transiting compact object and characterize the transiter (Sect. 4).

\section{$Kepler$ light curve of KOI-13}

{The visual double star KOI-13 (BD+46 2629, CCDM J19079+4652AB, Fig. 1 top panel) was discovered by Aitken (1904), who determined a separation of 1\farcs0 along a position angle of 281$^\circ$. The companions have apparent magnitudes $V_A$=9\fm9, $V_B$=10\fm2 (CCDM datalog, Dommanget and Nys, 1994). The position of the components remained constant since 1904.} $Kepler$ Short Cadence (SC) Q2 light curve of KOI-13 was taken in the integrated light from both components (Fig. 1, upper middle panel), and it shows a significantly asymmetric transit shape. This distortion can be easily perceived by eye. In Fig. 1, we plotted the folded Short Cadence light curve (epoch=JD 2454953.56498, period=1\fd7635892, Borucki et al. 2011) with a grid highlighting the most important asymmetries. The lower grid line intersects the light curve at a relative flux of 0.9956, and the orbital phase of the bottom of the transit is marked. Evidently, the floor of the light curve is shifted towards the ingress phase. The flux level of 0.9956 is observed at an orbital phase of $-$0.019 before the minimum, and 0.015 after the minimum, which is a severe light curve distortion. Also, the second contact, which is indicated by a sudden break of the light curve, occurs at 0.9964 flux in the ingress phase (this is pointed by the grid line); while the third contact happens at a slightly increased relative flux, 0.99665; which means 7\%{} difference in the eclipsed light.

Due to these distortions, one can hardly fit any of the widespread symmetric analytical models to the transit curve. The distortions can be checked against a non-parametric template, which is symmetrical in shape, but fits well the observations in other sense. Such a light curve template can be constructed by symmetrization. We constructed it from a spline fit through the  phase-folded light curve superimposed with its mirror inversion in phase. The residual curve is plotted in the lower middle panel of Fig. 1. The residuals contain about 1/40 of the total light variation.

KOI-13.01 can also be detected in {secondary eclipse} (Fig. 1, bottom panel), with an eclipse depth of 0.00012$\pm$0.00001 and eclipse duration ($t_1$ to $t_4$) of 3\fh0$\pm$0\fh2, while its mid-time occurs at 0.5004$\pm0.0004$. There is no hint for an eccentric orbit of KOI-13.01, {excluding} the eccentricity as the origin of transit shape distortions. The variation of the reflected light is observed in the out-of-transit phases, and it is also asymmetric in shape.
The models of Barnes (2009) predict exactly this kind of distortion in the transit light curve if KOI-13.01 ingresses near the hot polar spot of the rapidly rotating star, and the surface is cooler at the end of the transit path (Fig. 2).

\section{Properties of KOI-13 A and B}

\subsection{KOI-13 is a CPM binary}

In the past century, several observations of KOI-13 were taken, including 6 observations in the Tycho-2 Catalog. {Our latest imaging (Fig. 1, top panel) on April 20, 2011 was done at the Piszk\'estet\H{o} Mountain Station of the Konkoly Observatory, with the 1m RCC telescope equipped with a VersArray 1300b NTE camera (Szab\'o et al. 2010). The seeing was 0\farcs7 during this observation, enabling the secure separation of the stellar images and an accurate photometry. We determined a separation of 1\farcs18$\pm$0\farcs03 along PA~$281^\circ\pm1^\circ$, measured a brightness difference of $\Delta V=$0\fm20$\pm$0\fm04 and almost identical colors, $\Delta(V-I)=$0\fm10$\pm$0\fm05.} 

There are four independent positional measurements from the past century, published in Aitken (1904), Dommanget and Nys (1994), H\o{}g et al. (2000), and Mason et al. (2001). It is evident that the position angle of KOI-13 A and B remained constant with a value of 281 within 2$^\circ$, with no variation in time. Separation values were determined between 1\farcs1 and 1\farcs 15, which is constant or may have been very slightly increasing. Interstellar reddening is significant towards KOI-13: despite both components have an A-type spectrum, Tycho (B$_T$-V$_T$) colors of components A and B are 0.228 and 0.256, respectively.

%\begin{table}
%\caption{Astrometry and photometry of KOI-13. References: $^1$Aitken 1904, $^2$Dommanget et al. 1994, $^3$Hog et al. 2000, $^4$Mason et al. 2001, $^5$This work.}
%\begin{tabular}{lllll}
%\hline
%\hline
%Sep.($^{\prime\prime}$) & PA & $\Delta$ mag& $\Delta$ color & Epoch \\
%\hline
%1.0 & 281 & 0.13 & -- & 1904$^1$ \\
%1.1 & 282 & 0.3 & -- & 1950$^2$ \\
%1.15(4) & 284(1) & $\Delta$V$_T$=0.13(5) & $\Delta_{B_T-V_T}$=0.03(5) & 1988.6$^3$\\
%1.1 & 279 & -- & -- & 2001$^4$ \\
%1.18(3) & 281(1) & $\Delta$V=0.22(4) & $\Delta_{V-R}$=0.06(6) & 2011.31$^5$\\
% &  &  & $\Delta_{V-I}$=0.10(5)  & 2011.31$^5$ \\
%\hline
%\end{tabular}
%\end{table}

Two proper motion determinations are available for both components. Tycho-2 Catalog gives a common proper motion for both components, $\mu_{\alpha}=1.5\pm$1.5~mas/yr, $\mu_{\delta}=-16.6\pm$1.4~mas/yr. Most recently, Kharchenko and Roeser (2009) determined two independent proper motions, $\mu_{\alpha}^A =-0.82\pm1.5$, $\mu_{\delta}^A =-16.30\pm1.4$, $\mu_{\alpha}^B =-1.50\pm1.5$, $\mu_{\delta}^B =-16.60\pm1.4$. This is also consistent with a common proper motion (CPM) within the errors. The direction of the proper motion is 185$^\circ\pm6^\circ$, which is almost {\it perpendicular} to the relative position of the components. Thus a significant difference in the velocity of proper motion would largely change the position angle. A variation of about 4--5 degrees in the position angle would have been surely detected within the past years. From the lack of this observation, we get an upper limit for the {\it difference} of the proper motions, which is $S\times\tan 4^\circ /106$~yr$=0.7$~mas/yr (here S=1100 mas, the separation of the components).

We checked how likely neighboring stars can reproduce the proper motion of KOI-13 and mimic a CPM binary by chance. We extracted data for 151 stars from the $Kepler$ field from Hipparcos or Tycho catalogs, with parallaxes between 1.78--2.27 mas. Their mean proper motion is $\mu_{\alpha}=0.97$mas/yr, $\mu_{\delta}=-2.8$ mas/yr with a large standard deviation (10.6 and 11.8, respectively). The ambiguity in the velocity of KOI-13 B is less than 0.7 mas/yr. Only 5\% of the stars in its vicinity exhibit a proper motion within this limit around the measured value of 16.6 mas/yr. The direction of PA~45$^\circ$/225$^\circ$ is preferred the most by $Kepler$ field stars; proper motion of KOI-13 is inclined to this angle. Only 1--2\%{} of the stars exhibit a proper motion towards PA 180$^\circ$--190$^\circ$; and we infer that less than 0.1\%{} of the stars exhibit similar proper motion vector than KOI-13. In conclusion, kinematic information confidently suggest that KOI-13 A and B form a binary system.

\subsection{The stellar components}

We fitted stellar model spectra to the A and B components making use of the available photometry and a spectroscopic observation.
We took one high-resolution (R$\approx$32,000) optical spectrum at the Tautenburg Observatory on April 19, 2011, using the 2m Alfred Jensch telescope and the Coud\'e{} echelle spectrograph. The exposure time was 30 min, yielding an S/N of about 125/pixel. Binarity was not seen during the observation because of the seeing, and the slit was centered on the photocenter. The observed spectrum therefore includes the light from both components. In addition, we derived (reddened) $(B-V)_A$=0.222 and $(B-V)_B$=0.227 from Tycho colors. We also made use of the 2MASS photometry of KOI-13, K=9\fm425 (the two components together), leading to a reddened $(V-K)_{A+B}=0.52$ and $(J-K)_{A+B}=0.041$. Some of these data (the spectrum and $V-K$ color) were derived from the composite light of both components, only the brightness and B-V were measured for each component separately. The derivation of stellar parameters was therefore a complex task.

First we fitted the average $T_{\rm eff}$, $\log g$ and $v\sin i$ of the composite spectrum, assuming solar metallicity. Fixing these parameters and varying metallicity, we determined the metallicity to be [Fe/H]=0.2, which we assumed for both stars. Then we calculated composite model spectra from atmosphere models with LLmodels program (Shulyak et al. 2004) in its most recent parallel version, and the SynthV program (Tsymbal 1996) to compute the grid of synthetic spectra on a cluster computer at TLS. Based on our lucky image photometry, we assumed 0.55 and 0.45 relative flux of the two individual components, and re-fitted radial velocity and $v\sin i$ values for each star in the region of metallic lines between 5000-5700~\AA{}. A rapid stellar rotation has been revealed for both components, with $(v \sin i)_A=70\pm10$~km/s and $(v\sin i)_B=65\pm10$~km/s. Because no significant difference was detected in the fitted radial velocities, we took the measured $-$7~km/s radial velocity for both components. $T_{\rm eff}$ and $\log g$ was then fitted using the entire 4700--5700 \AA wavelength range, including the $H\beta$ region. A $\log g$ of 3.9 and 4.0 was resulted for the two components, and we experienced slightly larger $\log g$ values for the fainter component. The final model spectrum fit is shown by the dashed (blue on-line) curve in the left panels of Fig. 3.

The measured value of $\log g$ suggests that both components are evolved turn-off stars. Thus, the determination of age, masses and reddening is possible from isochrones. For this task, we used Padova isochrones (Bertelli et al. 2008) with Z=0.016 metallicity. Since the turn-off is a good indicator of age, first we fitted the age of the appropriate isochrone in the $T_{\rm eff}$--$\log g$ plane, which is unaffected by interstellar reddening. The best-fit value was $\log t=8.85$ (700 Myr). Then the initial masses of both components, $E(B-V)$, $E({V-K})=2.69\times E({B-V})$ (Larson and Whittet 2005) and $A_V=3.1\times E({B-V})$ were fitted simultaneously, observing the difference between the model-predicted and the measured $T_{\rm eff}$, $\log g$, and dereddened colors. Finally, we determined two A-type stars with $L=30.5L_\sun$ and 23$L_\sun$, both evolving to subgiant state and rotating rapidly (Fig. 3). The fitted parameters and their confidence regions are summarized in Table 1.

\subsection{The host star of the transiter: KOI-13 A}

KOI-13 is a double star, but it is unresolved in the $Kepler$ images, because of the coarse image sampling of 3\farcs98/pixel. The host star of KOI-13.01 has to be identified before further investigation. With two independent tests, we found that 
that the brighter star, KOI-13 A, hosts the transiting component.

We did the pixel-level photometry based on the publicly available Kepler target pixel files\footnote{MAST: http://archive.stsci.edu/kepler/}.
Few pixels in the central column of KOI-13 image are saturated, but the neighboring 18--20 pixels can be evaluated in each segment in the wings of KOI-13 images. We segmented the vicinity of KOI-13 into 4 quarters, along the pixel's X-Y direction. Because the $Kepler$ field is rotated, KOI-13 A is in the left segment, and the largest light variation of KOI-13 A is expected to be detected here. The transit depth was measured to be 7.1~mmag in the left image quarter, while the depth was 3.8--5.7~mmag elsewhere. The errors of these values are less than 0.1 mmag. 

A transit occurring on 17/18 April was observed from the Piszk\'estet\H{o} Station of the Konkoly Observatory. Lucky imaging technique was applied with the {VersArray 1300b NTE camera on the 1m RCC telescope}. In total, 4694 images were taken with 1 sec exposure times. The seeing was about 1\farcs5, and the two components could have been resolved on $\approx$15\%{} of all images. Despite the mediocre seeing, there was chance to do photometry with image synthesis technique: the images were fitted with a superposition of two PSFs.  The PSF was defined for each individual images based on the observation of two bright stars in the vicinity of KOI-13. The relative position of KOI-13 A and B was kept fixed, and their joint position and the fluxes were allowed to vary. Linear resampling was applied to allow sub-pixel accuracy in the position of the model PSFs.

In the lower panel of Fig. 4, we plot the result of this photometry. Small dots show the binned median of the individual flux ratios in a bin size of 27 data points. The points show the same with a bin size of 162, while error bars denote the error of the mean (i.e. standard deviation of all 162 points/$\sqrt{162}$). For comparison, the predicted brightness variation is also plotted, assuming that KOI-13 A diminishes. The general similarity verifies that KOI-13 A is the one that gets fainter and thus it is the host star of KOI-13.01.

\subsection{KOI-13.01 is a brown dwarf}

The transit light curve has to be corrected for the light from KOI-13 B. Assuming $\Delta$V$\approx$0\fm2, KOI-13 B contains 45\%{} of the observed light, and the appropriate transit light curve can be derived by multiplying the excess flux by $\approx$1.818. The measured depth of the transit is 4600 ppm before applying the correction, thus the corrected depth is 8400 ppm. We fitted a parametric planet model (Mandel and Agol 2002) to the reconstructed light curve via the relative size of the planet, $r_p/r_*$, the impact parameter $b$ and the transit duration $D$, assuming quadratic limb darkening with $G_1=0.268$, $G_2=0.312$ (Claret, 2003). The resulting values are $r_p/r_*=0.0884\pm0.0027$, $b=0.75$, $D=2.824\pm0.002$~h. Here the errors are conservative estimates, assuming 3\%{} ambiguity in the correction for the second light. Despite the fit is not perfect due to the prominent distortion in the light curve, the relative size of KOI-13.01 is accurate enough to determine the radius of the companion, which results to be 2.2$\pm$0.1~$R_J$. There is no example for such large planets among the known exoplanets (their upper size limit is about 1.7 R$_J$, e.g. Szab\'o and Kiss, 2011).
A highly irradiated, very young ($\approx$10 Myr), 0.1 M$_J$ mass and coreless planet may experience this size for few
million years (Fortney et al. 2007), but the derived age of KOI-13 (700 Myr) excludes such a young age. Thus, we conclude that KOI-13.01 is most likely a brown dwarf, or possibly at the very low-end of the red dwarf stars (Demory et al. 2009).

{
This radius is too large for a brown dwarf which is not irradiated (Baraffe et al. 2003). A red dwarf may be suggested by the observed deep secondary eclipse. The depth of the secondary eclipse is $\approx$1/130 of the transit, which implies a $\approx$3150~K effective temperature of the companion, based on our model fit in $Kepler$ bandpass. This temperature is evidently too low: a red dwarf could not exhibit the observed radius unless its temperature is 3500~K (0.2 M$_\sun$, 700~Myr age).

To estimate the equilibrium temperature of the companion, $T_{\rm eff}^p$, we assume uniform heat redistribution, $T_{\rm eff}^p = {L^* \over 16 a^2 \pi \sigma}$. Here $L^*$ is the luminosity of the star and $a$ is the semi-major axis of the companion. By expressing the luminosity of the star with its $R^*$ radius and $T^*$ temperature, and making use of Kepler's third law, we finally get the followings:}

\begin{equation}
{T_{\rm eff}^p \over T^*} = \sqrt{R^* \over 2 a} \ \ \equiv \ \ \sqrt{R^*/R_\sun \over 430 \ M^*/M_\sun}  \ \sqrt[3]{1\over P_{\rm [yr]}}.
\end{equation}

{
The orbital period $P$ is expressed in years. The first part takes the $R^* / a$ term of a light curve fit as input. The right side is identical, but expressed via stellar model parameters. Substituting the parameters derived for KOI-13 A  we get $T_{\rm eff}^p\approx$2700~K. The estimated effective temperature exceeds the equilibrium temperature by about 17\%{}. However,  taken an uneven temperature distribution into account, a day-side temperature of $\approx$3000--3100 K is plausible without internal heat production. This makes plausible that KOI-13.01 is a brown dwarf indeed, and its radius may be inflated simply because of the highly irradiated environment.
}

\section{Discussion}

KOI-13.01 is the first example for a detection of light curve distortion due to the rapid rotation of the host star. The asymmetry also implies orbital obliquity for KOI-13.01. The distortion of the transit curve is small and a unique global solution is very difficult to find, if possible at all (i.e. fitting the stellar rotation and the surface brightness distribution). Nevertheless, it is evident that the orbit is significantly oblique. Oblique orbits have been found in several exoplanetary systems, offen considered as diagnostics of an orbital evolution affected by scattering processes (cf. refs. in Sect. 1). In addition to this picture, KOI-13 A has a distant stellar companion, KOI-13 B, and obviously, this system must have had a fascinating dynamical history.

KOI-13.01 is the prototype of a ``hot'' brown dwarf of an early-type host star, which was unprecedented before. Host stars of some exoplanets (HAT-P-7, KOI-428, WASP-33, OGLE-2) have masses between $1.47 M_\sun < M_* < 1.52 M_\sun$, predominantly F-type stars, while WASP-33 is an A5-type star (Guenther et al. 2011). In the radial velocity data, there are 24 companions belonging to stars with $M_*>1.8 M_\sun$, and some of these objects are very probably brown dwarfs. However, these companions are not ``hot'', their orbital periods span from 127 days (HD~102272) to 1191 days ($\kappa$ CrB). 
It is interesting to recall the suggestion (Lovis and Mayor 2007, Bowler et al. 2010) that more massive stars tend to produce more massive planets, probably because the mass of the giant planets is scaled by the mass of the protoplanetary disk. {The discovery of a close-in late-type dwarf} around a massive, A-type star tends to confirm this hypothesis.

The configuration we have unraveled in KOI-13 must be rare. Searching for similar asymmetries, we tested all 474 $Kepler$ candidates that had a vetting flag of 2 (i.e. passed all tests for planet nature) and which had a single exoplanet candidate companion. KOI-13 is the only clear example for such a light curve distortion. Fortunately, KOI-13 is bright enough for further observations with powerful instruments. We note that the method of symmetrized template may be too crude for the detection of fine distortions, and asymmetric analytical templates may perform better. Development of such an analytical method is in progress.

\section*{Acknowledgments}

This project has been supported by 
the Hungarian OTKA Grants K76816, K81421, K83790 and MB08C 81013, and the ``Lend\"ulet'' Young
Researchers' Program of the Hungarian Academy of Sciences.

~

~

\newpage

\begin{table}
\caption{Fitted stellar parameters of KOI-13 A and B}
\begin{tabular}{lll}
\hline
\hline
 & comp. A & comp. B\\
\hline
$\log T_{\rm eff}$ & 3.930$\pm0.020$ & 3.915$\pm$0.020 \\
$\log g$ & 3.9 $\pm 0.1$ & 4.0$\pm$0.1\\
$[Fe/H]$ & 0.2 & 0.2 \\
$v \sin i$ & 65$\pm$ 10 & 70$\pm$10\\
$B-V$ & 0.09 $\pm$ 0.04 & 0.11 $\pm$ 0.04\\
%$V-K$ & 0.17$\pm$ 0.1 & 0.19$\pm$0.1 \\
$M/M_\sun$ & 2.05 & 1.95 \\
$R/R_\sun$ & 2.55 & 2.38 \\
$M_V$ (mag) & 1.06 & 1.35\\
$A_V$ (mag) & 0.34 & 0.34 \\
$\log t$ & 8.85 $\pm$0.1& 8.85$\pm$0.1\\
Dist. & 500 pc & 500 pc \\
\hline
\end{tabular}
\end{table}

\newpage

\begin{figure}
~~~~~~~~~~~\includegraphics[width=9cm]{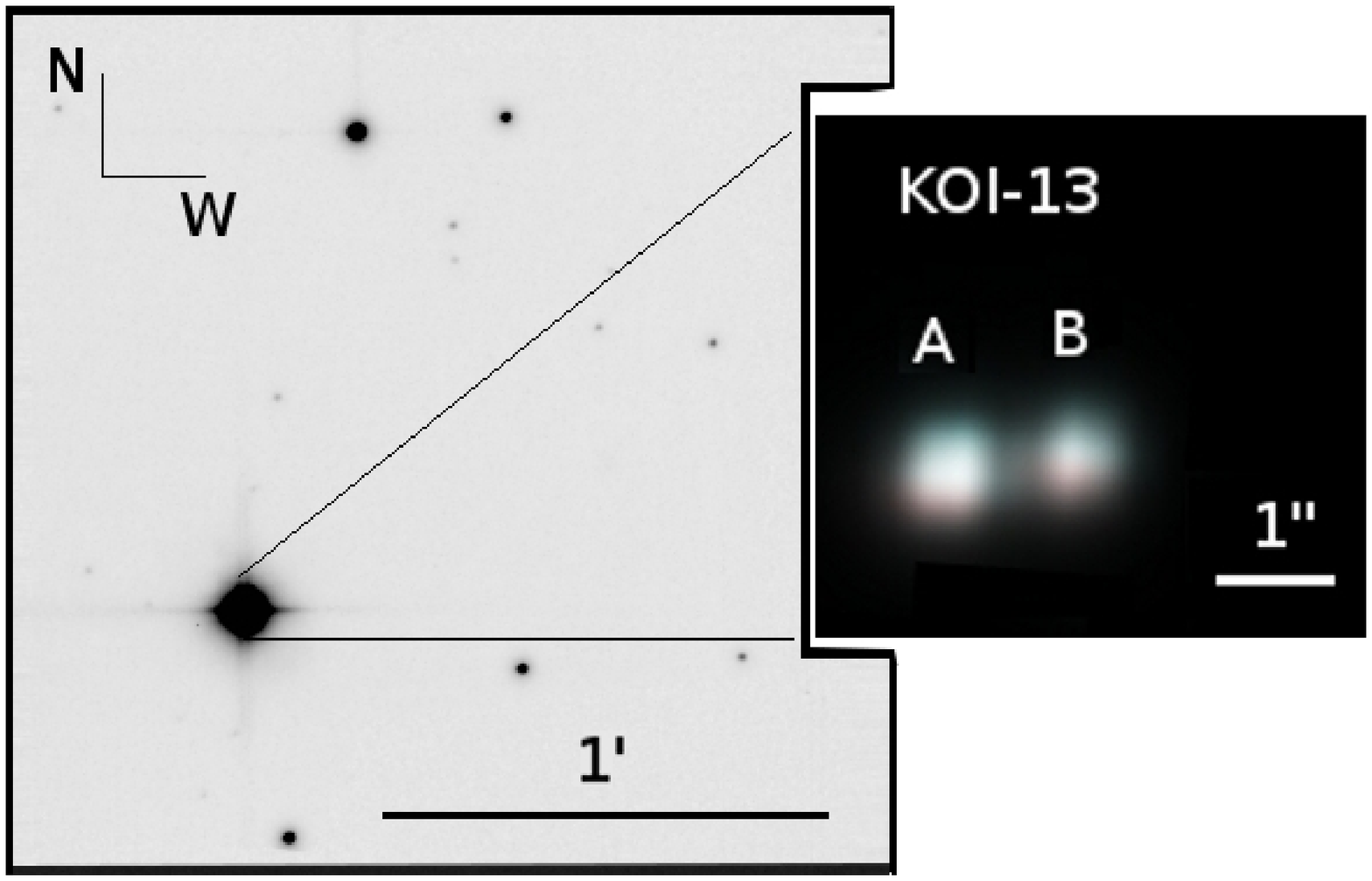}
\centering\includegraphics[width=13cm]{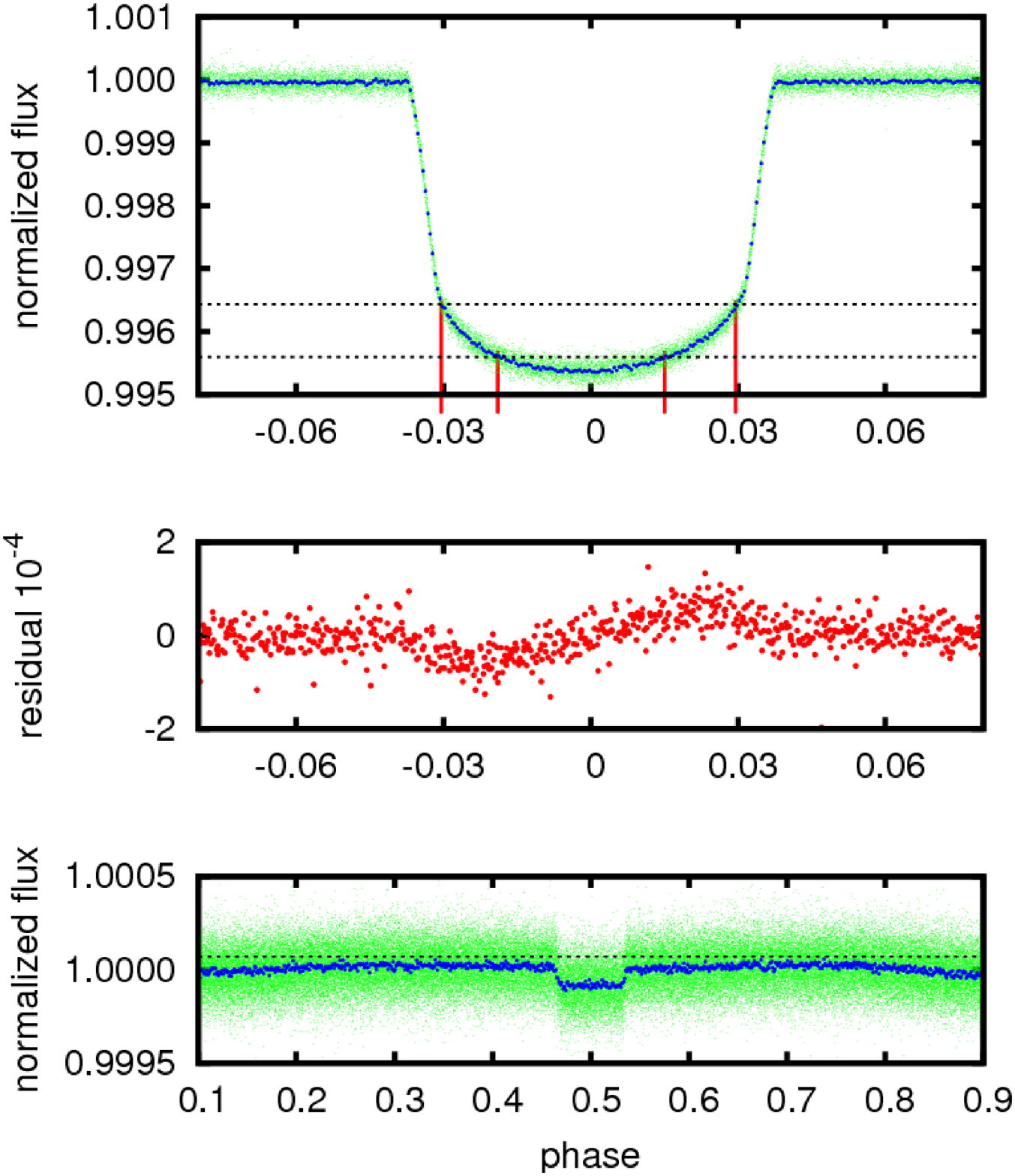}
\caption{Top: the vicinity of KOI-13 (left panel) and its false color (V,I) image with lucky imaging
(1m RCC telescope, Konkoly Obs., 2011 April 20). Upper middle: folded $Kepler$ light curve of KOI-13 (A and B components together; epoch=JD 2454953.56498, period=1\fd7635892, Borucki et al. 2011). Lower middle:  the residuals of the transit to a symmetric template. Bottom: the out-of-transit phase and the eclipse. Note the grid lines which emphasize the distortions in the light curve.}
\end{figure}

\newpage

\begin{figure}
\centering\includegraphics[width=9cm]{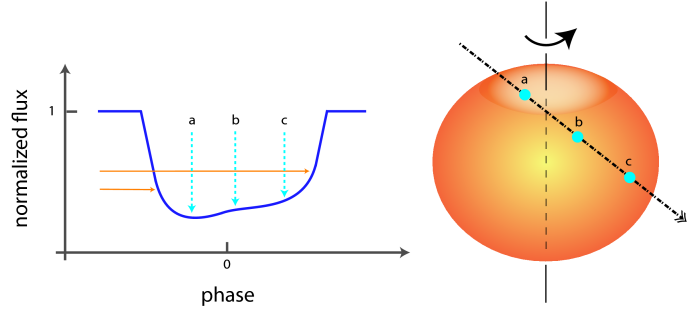}
\caption{The observed distortion is qualitatively consistent with stellar temperature gradient due to gravity darkening caused by the rapid rotation of the host star. This schematic figure is based on the Barnes (2009) {models}.}
\end{figure}

\newpage

\begin{figure*}
\centering\includegraphics[angle=270,width=15cm]{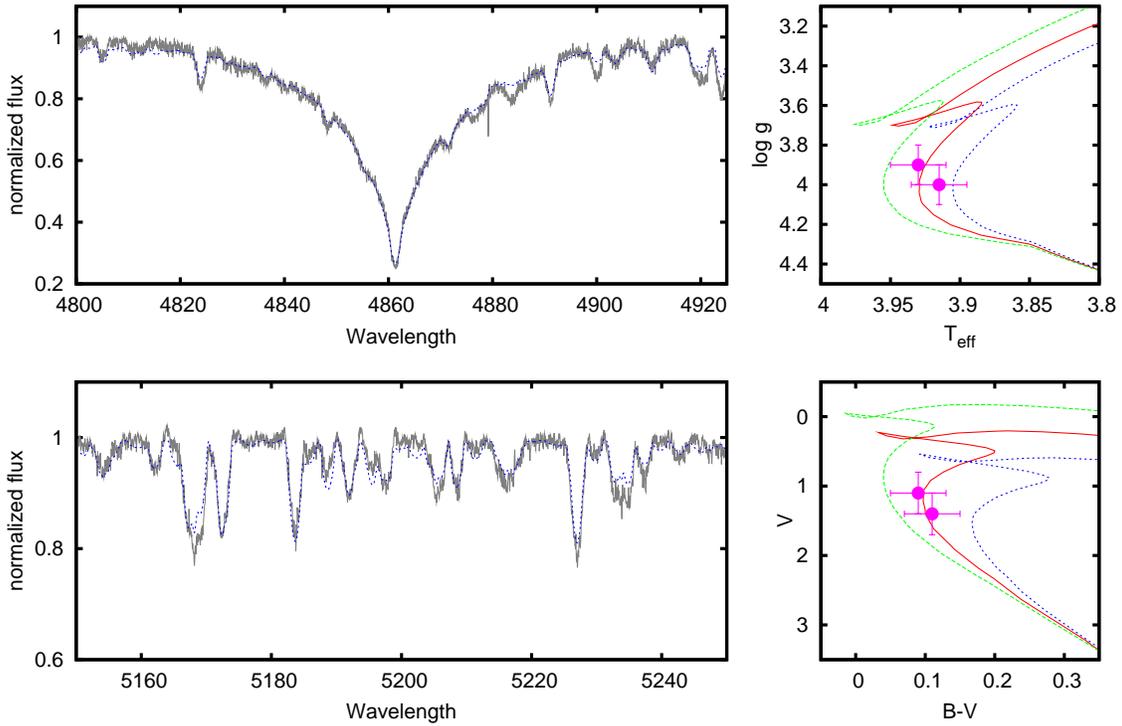}
\caption{Determination of stellar parameters of KOI-13. Left panels: details from the observed spectrum (gray) with the best-fit model (dashed, blue on-line). Right panels: fitting of age, mass, $B-V$ and luminosity with isochrones. Lines show the probed Padova isochrones with $\log t$=8.75, 8.85, 8.95, respectively.}
\end{figure*}

\newpage

\begin{figure}
\centering\includegraphics[width=9cm]{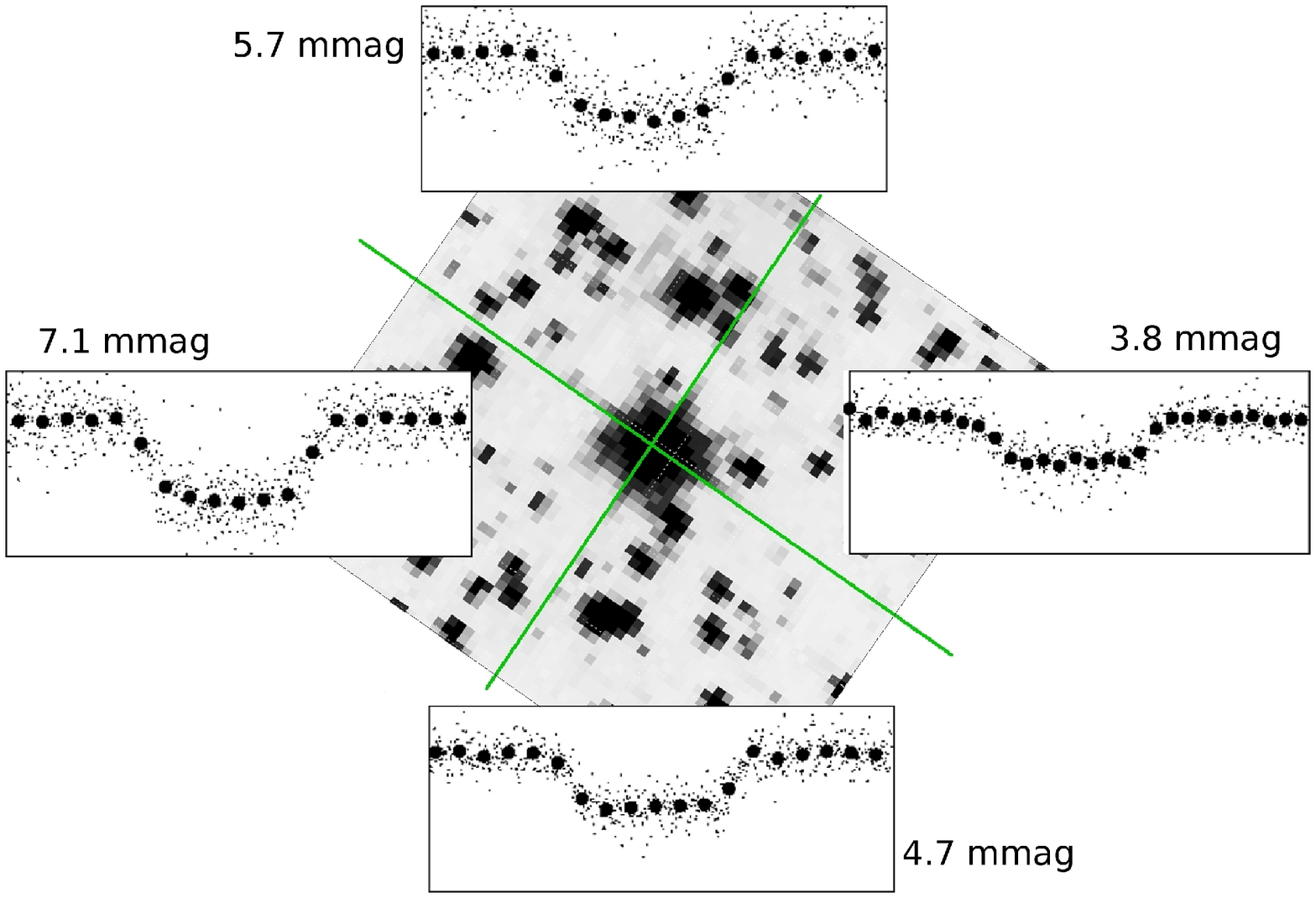}

~

\centering\includegraphics[bb=126 246 485 485, width=6cm]{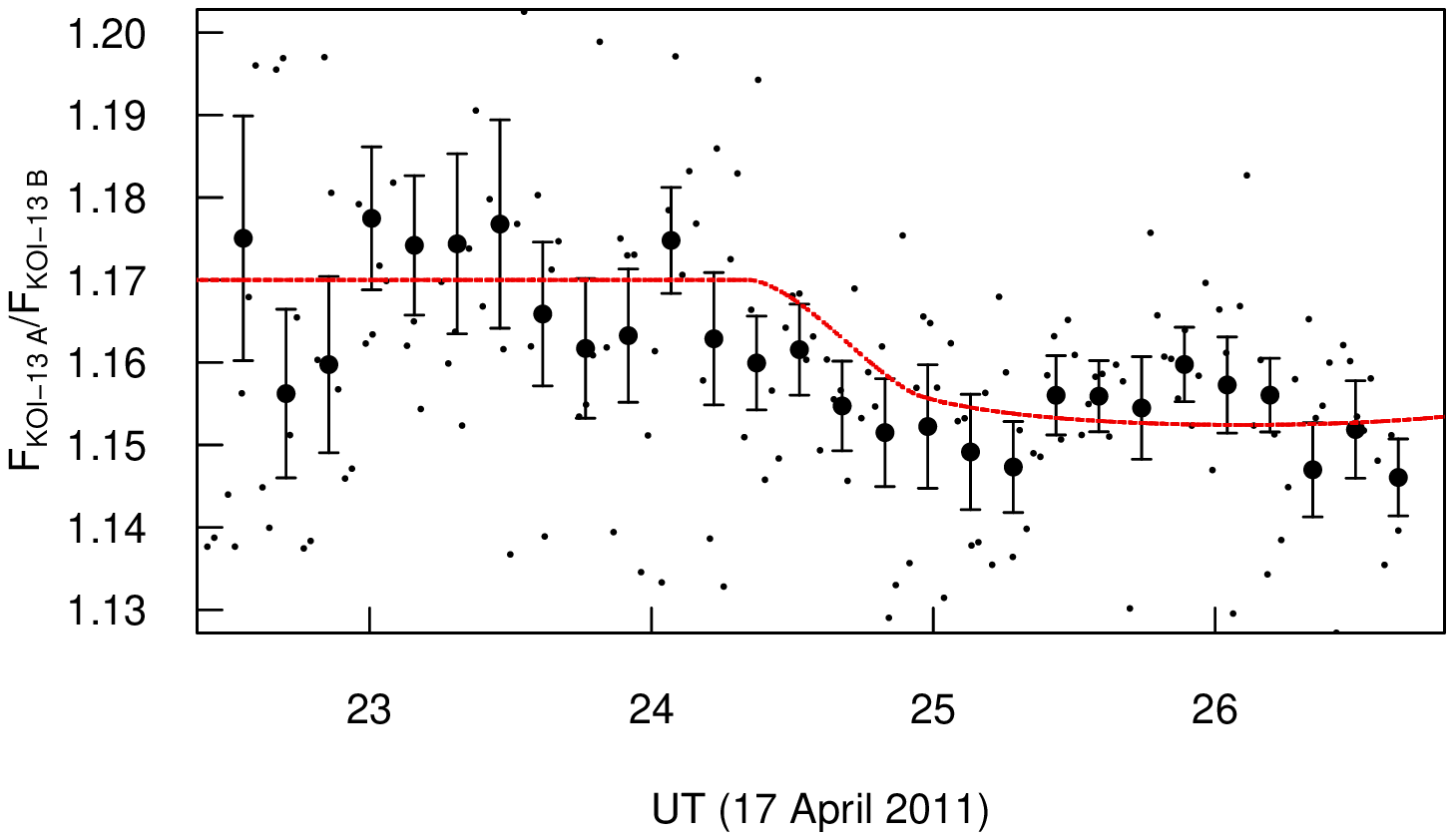}
\caption{Top: $Kepler$ light curves from pixel-level photometry of unsaturated regions in the vicinity of KOI-13. The largest amplitude is observed to the nearest of KOI-13 A, the host star of KOI-13.01. Below: Konkoly light curve of KOI-13 A in respect to KOI-13 B on 17 April, 2011. For comparison, a prediction is plotted if KOI-13 A is the host of KOI-13.01.}
\end{figure}

\end{document}